\documentstyle[12pt,aaspp4]{article}
\begin{document}
\title{A Search for Stars of Very Low Metal Abundance.  IV.  
$uvbyCa$ Observations of Metal-Weak Candidates from the Northern HK
Survey}
\author{Barbara J. Anthony-Twarog}
\affil{University of Kansas}
\author{Ata Sarajedini}
\affil{Wesleyan University}
\author{Bruce A. Twarog}
\affil{University of Kansas}
\author{Timothy C. Beers}
\affil{Michigan State University}
\authoraddr{Department of Physics and Astronomy, University of Kansas, Lawrence, KS 66045-2151}
\authoraddr{Wesleyan University,Astronomy Dept., Middletown, CT  06459-0123}
\authoraddr{Michigan State University, Dept. of Physics and Astronomy, E. Lansing, MI 48824}
\affil{Electronic mail: anthony@kubarb.phsx.ukans.edu,ata@astro.wesleyan.edu,
twarog@kustar.phsx.ukans.edu, beers@pa.msu.edu}

\begin{abstract}
CCD photometry on the $uvbyCa$ system has been obtained for 521 candidate
metal-poor stars from the northern hemisphere HK survey of Beers and
colleagues.   Reddening corrections, classifications of stars by luminosity
class, and the derivation of metal abundances based on Str\"omgren indices
$m_1$ and $hk$ are described and presented, along with preliminary 
spectroscopic estimates of abundance.  A number of extremely metal-poor stars
with [Fe/H]$ \leq -2.50$ are identified.

\end{abstract}

\section{Introduction}

A common goal of Galactic structure studies is to understand the events and
processes that have characterized the formation and ongoing evolution of the
Milky Way.  Surviving relics of early star formation in the Galaxy offer
particularly rich rewards, but these rewards only come as a result of
relatively large observational effort.  Clues from the first epochs of star
formation are scarce and challenging to decode, but they can be explored via
examination of the stars of the halo field and in the populous globular
clusters.

Over the past few decades, researchers have probed the spatial, kinematic and
chemical distribution of the stars formed over the halo's history.  Since the
seminal explorations of Eggen, Lynden-Bell, \& Sandage (1962)\markcite{els},
the amount and quality of available data has grown to the point that a rather
more complicated picture of the formation of the Galaxy than that based on a
simple, rapid, collapse of a pre-Galactic spheroid of gas now applies (see,
e.g., Chiba, Beers, \& Yoshii 2000\markcite{chi}).

While the pioneering work of Eggen et al. focused on the properties of a
kinematically selected sample of nearby stars, one must concede
that, in many ways, Galactic globular clusters are ideally suited for mapping
out this history.  The advantages of the globular clusters are well documented
in the literature.  Each globular cluster, comprised of $10^5$ to $10^6$ stars
which share a common age, distance, and (usually) chemical composition permits
determination of these fundamental properties with minimized statistical
errors.  Furthermore, globular clusters are readily observable at distances
from the Galactic center out to many tens of kiloparsecs.  Relative ages,
precise to 1 Gyr or better, have been reported for halo clusters (Sarajedini,
Chaboyer, \& Demarque 1997\markcite{sd}; Buonanno et al. 1998\markcite{bcp}).
The ``merger model'' for halo formation (Searle \& Zinn 1978\markcite{sz}), as
well as the existence of a thick disk population of Galactic globulars is
supported by the kinematic and spatial distribution within the globular cluster
population (Zinn 1993\markcite{zi}; Da Costa \& Armandroff 1995\markcite{dca}).

There remains, however, one devastating weakness in the effort to reconstruct
the history of Galactic evolution with globular cluster data alone -- there
exist no more than $\sim 200$ globulars in the Milky Way.  Furthermore, there
is no globular cluster located within one kpc of the Sun, thus restricting
high-resolution stellar studies to the brighter giants, and requiring
significant commitments of large telescope time.  In addition, there are no
known clusters with abundances below [Fe/H]$ = -2.5$, in sharp contrast with
the field star sample (Laird et al.  1988\markcite{lcl}).  The exceptionally
metal-poor field stars, according to all models of Galactic nucleosynthesis,
should probe the first generations of star formation within the protogalactic
cloud.  Thus, if one seeks to study the {\it earliest} phase of Galactic
evolution, field stars offer the best option.

The picture that is now emerging from examination of nearly primordial stars is
fascinating but puzzling.  Abundances of the light elements seem well-described
by the standard model of Big Bang nucleosynthesis (Olive \& Steigman
1995\markcite{sch}; Ryan et al. 2000\markcite{rya}).  For the heavier elements,
stars with [Fe/H] $> -2.5$ have, within broad tolerances, well-defined patterns
of abundance ratios at a given [Fe/H], presumably reflecting their origin
within a reasonably well-mixed interstellar medium by the time these stars were
formed (Wheeler, Sneden, \& Truran 1989 \markcite{wst}; Edvardsson et al. 1993
\markcite{e93}; McWilliam et al. 1995 \markcite{mcw}; Ryan, Norris, \& Beers
1996\markcite{rnb}).  Stars with [Fe/H] $< -2.5$ (McWilliam et al. 1995
\markcite{mcw}; Ryan et al. 1996\markcite{rnb}) exhibit a more chaotic
distribution in a number of their elemental ratios, which may be the earmark of
an poorly-mixed early Galaxy.  Recent attempts to model this apparently complex
early history have centered on supernova-induced star formation and chemical
evolution in individual proto-Galactic clouds (Tsujimoto, Shigeyama, \& Yoshii
1999\markcite{tsy} and references therein).

Continued progress in this area is assured only if the sample of extremely
metal-poor stars is expanded -- by no means an easy task given that such stars
make up a tiny fraction of the already sparsely populated halo component among
stars in the solar neighborhood.  Early attempts based upon objective-prism
spectroscopy (Bond 1980\markcite{b1}; Bidelman \& MacConnell
1973\markcite{bm73}) generated only a handful of stars which proved to exhibit
metal abundances near [Fe/H] $\sim -3.0$.  The most effective survey technique
to date has been that developed by Beers, Preston, \& Shectman (1985, hereafter
BPS)\markcite{bps} in which the wide-field survey capability of a Schmidt
telescope, used with an objective prism, is combined with a filter isolating
the spectral region near the ultraviolet lines of singly ionized calcium.  The
strong H and K resonance lines of CaII should be visible in even the most
metal-poor stars, so stars with nearly continuous spectra in this region are
good candidates for extreme metal deficiency.  Unfortunately, blended stellar
images, high temperatures, and low surface gravity can also produce weak H and
K lines, thus one requires medium-resolution spectroscopy and/or photometric
follow-up to isolate the best candidates.  For an example of how severe this
winnowing process can be, one need only refer to the initial efforts of BPS.
In their first attempt, plate material covering 1940 square degrees of sky was
used to generate a list of 2000 candidate stars; from an initial subset of 450
candidates, roughly 30\% proved to have [Fe/H] below --2.0, with only a few
stars of extreme ([Fe/H] $< -3.5$) metal paucity being detected.  Note,
however, that prior to this study, only one star was known with an estimated
metallicity near --4.0 and virtually every major analysis of the primordial
halo over the past decade has included stars from the HK survey of Beers and
colleagues.  Beers (2000)\markcite{b20} discusses how the ``effective yield''
of such survey efforts can be improved, and points out that one can roughly
double the efficiency of searches for stars with [Fe/H] $< -2.0$ by conducting
a photometric pre-selection of candidate stars.

The HK survey has now been expanded to include a total of some 7000 square
degrees in the both the northern and southern hemispheres (see Beers
1999\markcite{b99} for a summary).  Medium-resolution spectroscopic follow-up
of the first portion of this expanded sample has been discussed by Beers et al.
(1992b, hereafter BPS II)\markcite{bps2}. Photometric data for subsets of the
long lists of candidates have been presented in Doinidis \& Beers (1990,1991)
\markcite{db1}\markcite{db2}, Preston, Shectman, \& Beers
(1991)\markcite{psb}, Beers et al. (1992a)\markcite{bpsd},
Schuster et al. (1996, 1999)\markcite{s0}\markcite{s99},
and Norris, Ryan, \& Beers (1999)\markcite{nrb}.

In this paper we analyze $uvbyCa$ CCD photometry for over 500 low metallicity
candidates from the HK survey, the majority of which were selected from
northern survey plates.  Section 2 describes the photometric system and the
detector; Section 3 describes the acquisition and processing of the CCD data.
In Section 4 the photometric classification and reduction to the standard
$uvbyCa$ system are discussed. In Section 5 we describe estimates of the
reddening corrections for these stars, and in Section 6 the derivation of
[Fe/H] from $m_1$ and $hk$ indices is considered, along with a comparison to
spectroscopically estimated abundances.  Finally, Section 7 presents a summary
of the principal results.

\section{The Instrumental System}

To estimate basic stellar parameters for our candidate metal-poor stars, we
have developed an extension of the basic intermediate-band photometric
system developed and described by Str\"omgren (1966)\markcite{st}.  The
original four $uvby$ bandpasses have full-widths of 160 to 300 \AA\ and central
wavelengths ranging from 3520 to 5500 \AA.\  The color index $(b-y)$ is
primarily a temperature index. Two other indices are color differences; 
$m_1 = (v-b) - (b-y)$ is designed to quantify the effect of line-blanketing in
the $v$ bandpass, and $c_1 = (u-v) - (v-b)$ provides an estimate of
surface gravity for spectral types A or later by measuring the change in
continuum slope on either side of the Balmer jump.  Magnitudes constructed from
the $y$ band are essentially indistinguishable from broad-band $V$
measurements.

Although originally designed for warm stars of near-solar abundance, the $uvby$
system has been successfully adapted and pushed to stars of lower metallicity
and cooler surface temperatures.  Bond (1980) \markcite{b1} presented one of
the earliest applications of the $uvby$ system to metal-poor red giants,
showing that metallicity and gravity estimates were obtainable with reasonable
accuracy.  Subsequent work by Anthony-Twarog \& Twarog (1994)\markcite{a3}
provided a calibration of [Fe/H] for giants based on their position in the
$m_1,(b-y)$ plane, and also presented methodology by which one might determine
foreground reddenings.  Application of $uvby$ photometry to very metal-poor
dwarfs is exemplified by the work of Schuster \& Nissen (1988)\markcite{sn88}.

The diminishing sensitivity of the $m_1$ index to changes in [Fe/H] for
metal-poor stars can be seen in the papers describing the calibration for F and
G dwarfs (Schuster \& Nissen 1989\markcite{s1}), and for late-type giants
(Anthony-Twarog \& Twarog 1994\markcite{a3}).  This limitation at low abundance
was the primary motivation for the extension of the Str\"omgren system in the
form of the $Ca$ filter introduced and described by Anthony-Twarog et al.
(1991)\markcite{a1}.  The $Ca$ filter has a full width of less than 100 \AA\
and includes the resonance lines of calcium in the near ultraviolet, lines
that remain reasonably strong even in quite metal-poor stars.  The $hk$ index
is constructed in a manner analogous to the $m_1$ index; $hk = (Ca-b) -
(b-y)$.  The  catalog of Twarog \& Anthony-Twarog (1995)\markcite{t1} lists
$hk$ indices, $V$  magnitudes, and $(b-y)$ colors for nearly 2000 stars.  A
more recent paper (Anthony-Twarog \& Twarog 1998\markcite{a2}) provides a
revised metallicity calibration based on $hk$ indices for metal-poor red
giants.

Following the model for the development of the $uvby$ photometry system, all of
the studies cited above  have been conducted with photoelectric detectors, with
the associated restrictions on areal coverage and survey depth.  Application of
CCD detectors to this and other photometric systems has broken through these
barriers.  In the course of the present project, we employed the CCDPHOT system
on the 0.9m telescope at Kitt Peak National Observatory.  CCDPHOT is a
real-time, IRAF/ICE-based CCD photometer described by Tody \& Davis
(1992)\markcite{td}, as well as in instrument manuals distributed by the
National Optical Astronomy Observatories (NOAO).

CCDPHOT extracts photometric data from processed data frames, preserving only
the photometric records for a limited number of stars from each frame.  As the
data are obtained in each filter, frames are corrected for overscan, bias, and
flat-fielding.  Aperture magnitudes are then extracted for up to three star
positions --  only the data relevant to the construction of aperture magnitudes
are retained.  At the end of a sequence of observations through the filters, or
a series of such sequences, instrumental magnitudes are constructed for the
targeted stars as observed in each filter, based on the cumulative flux and sky
measurements.

During our multi-year campaign, the CCDPHOT system was outfitted with
two different CCD detectors.  Both chips were small (512 by 512 pixels) devices
with an image scale of 0.77 arcsec/pixel; the first one was designated T5HA; it
was replaced in late 1995 with a similar device called T5HC. The readout noise
for both detectors was $\sim 4$ electrons/pixel. The primary difference between
these two CCDs was that the quantum efficiency (QE) of T5HA declined steeply
blueward of 4000 \AA\, affecting measurements in the near ultraviolet,  whereas
the QE of T5HC was significantly flatter in this wavelength regime (see the
comparison provided by Sarajedini et al. 1996\markcite{sc}).

\section{Acquisition and Processing of Data}

Measurements of photometric indices have both ``internal'' and ``external''
errors associated with them.  With respect to internal error, or precision, there
are the usual limitations imposed by photon statistics and by the stability of
photometric conditions.  Patient acquisition of sufficient photons to assign
$\sim 2$\% precision for a given magnitude measurement ($ > 2500$ photons from
the star alone) was achievable for nearly all the stars in the present sample;
the more stringent limitation was that of stability or repeatability of
photometric measurements.  Our standard observing procedure was to run through
the sequence of filters for each star at least twice;  additional integrations
were often added to achieve adequate photon statistics in the narrower or bluer
filters.  At the end of each night, the CCDPHOT software produces an
instrumental magnitude for each filter and each star based on the accumulated
photons in each bandpass.  However, we modified the initial processing
steps to examine the separate measurements in each sequence of observations for
each star.  Intercomparison of the measured magnitudes in each bandpass
provides an estimate of the precision which was achieved, as estimated by the
standard deviation of all repeat observations for a given star.  In cases where
only a single night's data is available, this estimate of precision is the only
one available.

Processing of each night's data yielded a list of instrumental magnitudes and
colors with errors that reflect the repeatability of the measurements.  Using a
few stars observed at different airmasses, extinction correction coefficients
were determined for most individual nights.  In some cases, extinction
coefficients were adopted from one night, and applied to other nights in the
same run.  The values of the extinction coefficients were within nominal limits
for Str\"omgren indices: $k_y$ ranged from 0.11 to 0.20, with typical values of
0.15 $\pm$ 0.02; $k_{b-y}$ ranged from 0.03 to 0.08; $k_{m1}$ ranged from 0.06
to 0.09;  $k_{c1}$ from 0.12 to 0.16; $k_{hk}$ from 0.11 to 0.14.

Where several consecutive nights of data were available, a master list of
instrumental, airmass-corrected magnitudes and indices was constructed.  This
procedure parallels our treatment of photoelectric photometry data; indices are
mapped by linear transformations, determined by stars in common among the
nights, to remove the minor instrumental differences between nights.   Average
indices for each star are constructed from the transformed data, then the
process is repeated in order to produce a master list of instrumental
magnitudes and indices.  Entries in Table 1 summarize the number of usable
nights during each assigned run, and the dispersion about a linear fit of each
night with respect to the master list of magnitudes and instrumental indices.

An identical procedure was carried out to place all of the runs on a common
master system, using the repeated observations of standard stars to provide
the link from one run to the next.  Indices from seven observing seasons 
between March 1995 through October 1997 were mapped and merged, using the
final October 1997 run as the instrumental template.   Table 2 summarizes 
the number of stars from each run in common with the master list, and 
the dispersions between the magnitudes and indices with respect 
to the master list.

For the $\sim $450 stars with single observations, the errors indicating
repeatability within a sequence of measurements may be summarized as follows:
the typical standard deviations for $V$, $(b-y)$, $m_1$, $c_1$ and $hk$
measurements are 0.017, 0.028, 0.045, 0.066 and 0.046, respectively.  If the
most problematic stars are excluded from these means, for example, by
clipping stars with standard deviations larger than three times these error
statistics, the standard deviations drop to 0.012, 0.020, 0.032, 0.043
and 0.034.  We conclude, therefore, that we have achieved 2\% or better
internal errors for $V$ and $(b-y)$, and perhaps only slightly worse than this
for the remaining indices.

In addition, there are 25 stars with measurements obtained during more than one
run throughout the period of the observational follow-up.  The standard
deviations describing the consistency of measurements for $V$, $(b-y)$, $m_1$,
$c_1$ and $hk$ for these stars are 0.011, 0.009, 0.016, 0.018 and 0.017.  These
errors are clearly smaller, which may indicate that our estimates of the
precision based on repeatability of single-star observations are somewhat
pessimistic.

\section{Classification and Reduction to Standard System}

Whatever the precision of the final instrumental photometry, the external error
is limited by the transformation to standard indices $V$, $(b-y)$, $m_1$, $c_1$
and $hk$.   Selection of a set of standard stars and indices is a critical
decision, especially for cooler stars, as the transformation equations are
distinguishably different for stars of different luminosity class.  An example
of the necessary distinction between giants and dwarfs is discussed
by Twarog \& Anthony-Twarog (1995)\markcite{t1}. Figure 1 of that
paper illustrates the divergence in calibrated $(b-y)$ colors for cooler dwarfs
and giants.  The $(b-y)$ colors in that catalog are tied to the Olsen
(1993)\markcite{o1} system for dwarfs and giants. The $V$ magnitudes, $(b-y)$
colors and $hk$ indices in the present study are tied to the 1995 catalog and
transformation system.

It is generally necessary to make some preliminary distinction between
luminosity classes before knowing which set of calibration equations to apply
to program stars.  Some classes of stars can be classified exclusively
through their position in the pseudo-HR diagram plane of the $c_1, (b-y)$
indices. Figure 1 of Pilachowski, Sneden \& Booth (1993)
\markcite{p1} shows this diagram for a variety of metal-poor star samples.  
The metal-poor giant branch is readily apparent; stars with $(b-y)\leq 0.5$ 
and $c_1 \geq 0.4$ are separable as horizontal branch stars, 
while very cool dwarfs and bluer main sequence stars are easy to distinguish. 
For moderately metal-poor stars ([Fe/H]$ \leq -1.5$), the position of the
giant branch in this plane is fairly insensitive to metallicity.
(Anthony-Twarog \& Twarog 1994)\markcite{a3}.

Separation into luminosity classes is less clear for F and G stars, where
metal-poor giants and subgiants may overlay the $c_1, (b-y)$ positions for
metal-rich dwarfs.  Fortunately, the additional information from the $m_1$
index makes this separation possible.  We used a rubric similar to Olsen's
(1993) scheme to classify stars into dwarf, giant and horizontal branch
categories by referring to $m_1$ indices as well as the $c_1, (b-y)$ plane.

With luminosity classifications in place, the linear equations that relate
instrumental indices to standard values were then determined.  The
Twarog \& Anthony-Twarog (1995)\markcite{t1} catalog is the source for
standardizing the $V$, $(b-y)$ and $hk$ indices.  The 84 cool giants and dwarfs
bluer than $(b-y) = 0.42$ defined one set of equations;  the linear relations
between instrumental and standard values have dispersions of 0.013, 0.010 and
0.017 magnitudes for $V$, $(b-y)$ and $hk$, respectively.  A modest color term
of $0.056 (b-y)_i$ is incorporated in the $V$ calibration by its addition to
the instrumental magnitude.  The 38 dwarfs with $(b-y) \geq 0.42$ from the same
catalog define the linear transformation for the cooler dwarfs, with
dispersions about these linear fits of 0.011, 0.008 and 0.016 magnitudes,
respectively.  These linear equations are similar, with an identical color-term
in the $V$ calibration, but include a substantially shallower slope for the
$(b-y)$ calibration for the dwarfs.

To maintain a link to photometric metallicity calibrations and reddening
determinations for cool giants, the $m_1$ and $c_1$ indices for cool giants in
Anthony-Twarog \& Twarog (1994)\markcite{a3} were used to transform these
indices for stars with instrumental $c_1$ indices indicating evolved stars.
Indices for 39 cool giants define the linear transformations, with dispersions
about the fitted lines of 0.017 magnitudes for $m_1$ and 0.023 magnitudes for
$c_1$.  A substantial color term is incorporated in the relationship fitting
standard $c_1$ indices to instrumental $c_i$ indices augmented by $0.30$
$(b-y)_i$.

Photometric indices from Schuster \& Nissen (1988)\markcite{sn88} were used to
transform the $m_1$ and $c_1$ indices for dwarfs.  A break in the linear
solutions was imposed at $(b-y) = 0.42$, with 49 standard stars bluer than this
criterion and 46 stars redder.  The dispersions about the linear fits of
instrumental to standard $m_1$ and $c_1$ indices were 0.019 and 0.020
magnitudes for the warmer dwarfs, and 0.020 and 0.020 magnitudes, respectively,
for the cooler dwarfs.

The 521 program stars were classified and calibrated through
the appropriate set of linear relations.  The resulting calibrated photometry
is presented in Table 3.  The first column lists the star name.  Columns (2)
and (3) list the 1950 epoch coordinates.  The Galactic longitude and and
latitude for each star are listed in columns (4) and (5).  Columns
(6)--(15) list the $V$ magnitudes, and photometric indices $(b-y)$, $m_1$,
$c_1$ and $hk$ and their associated formal errors, respectively.  Column (16)
indicates the number of photometric measurements made for each star.

Entries in Table 3 are, on the whole, restricted to include only those with
indices within the index limits of the standard stars.  These limits are
approximately: $(b-y)$ between 0.0 and 1.0; $m_1$ between
0.0 and 0.7; $c_1$ between $-0.05$ and 0.8; $hk$ between 0.3 and 1.7.  
Data for a
few stars with exceptions to these limits are noted in the table by indices
with a colon following the index value.  Photometric index entries do not
appear if the estimate errors exceed 0.1 for $V$ or $(b-y)$, with an error
threshold of 0.2 for the remaining three indices.  Figure 1 summarizes the
photometric errors for stars listed in Table 3.

A number of plates in the HK survey overlap with one another, hence there are
several instances of candidates with multiple names.  Table 4 summarizes
the multiple identifications for the HK survey stars in our present sample.

We have been able to compare our photometry with the results from a number of
other studies which included northern stars from the HK survey.  There are 31
stars in common with the list published by BPS II and with photometry in Preston
et al. (1991)\markcite{psb} and Beers et al. (1992a)\markcite{bpsd}.  We
excluded three stars with highly discrepant magnitudes -- CS 22949-0026, CS
22949-0025 and CS 22189-008 -- and found a dispersion of about 0.04 magnitudes
about a unit-slope relation between their $V$ magnitudes and ours, with no
significant offset. We were able to make a wider comparison of $V$, $(b-y)$,
$m_1$ and $c_1$ but for a very small sample, using five stars in common with
the primarily metal-poor turnoff sample studied by Schuster et al.
(1996)\markcite{s0}. Except for one highly discrepant $c_1$ value, that
comparison also suggests that no significant offset in any of the indices $V$,
$(b-y)$, $m_1$ or $c_1$, exists but that a dispersion of 0.04 magnitudes
pertains for all indices, rather larger than the scatter implied by our own
comparisons of stars observed over more than one season.

Norris et al. (1999) \markcite{nrb}  have published $UBV$ photometry for a
number of stars in the northern HK survey, 133 of which are found in the
present sample.  Once again, a substantial number have magnitudes sufficiently
discrepant to suggest that we observed different stars.  A comparison of $V$
magnitudes for the remaining 117 stars suggests a dispersion of $\sim 0.04$
magnitudes, with no significant offset or deviation from unit-slope.  A new
sample of Str\"omgren photometry for HK survey stars has been completed by
Schuster et al.  (1999)\markcite{s99}.  Comparison of 32 stars in common with
their published sample indicates the following differences (in the sense
Schuster et al. $-$ KPNO): for $V$, $0.014 \pm 0.024$, for $(b-y)$, $-0.002 \pm
0.022$, for $m_1$, $0.003 \pm 0.042$ and for $c_1$, $0.001 \pm 0.056$. This is
our largest multi-index comparison with an external source of photometry; the
errors generally confirm the less optimistic errors discussed in Section 3 that
were based on the repeatability of flux measurements.  We indicate the stars in
Table 3 which may be mis-identifications, on the basis of the above
comparisons, with an asterisk next to the star names.

Approximately 100 of our program stars have color indices bluer than $(b-y) =
0.25$ or redder than $(b-y) = 0.65$, placing these stars outside any available
photometric metallicity calibrations.  No further analysis has been attempted
for these stars, with the exception of an examination of the bluest stars for
potential white dwarf candidates.  Five stars have $(b-y)$ and $(u-b)$ colors
similar to DA white dwarfs in the compilation of Fontaine et al. (1985).
\markcite{f1}  These stars are BS 16090-0038, BS 16468-0013, CS 22171-0018,
CS 22898-0026 and CS 31067-0025.

\section{Reddening Corrections}

Correcting for interstellar extinction poses a challenging intermediate step on
the path toward photometric abundance estimates.  One consequence of applying a
significant reddening correction might be the re-classification of a star from
an evolved, cool star to a warmer dwarf or horizontal-branch star.  Equally
important, reddening-corrected colors are essential for placing a star
correctly in a photometric plane defined by indices influenced by chemical
composition as well as stellar temperature.  Photometric abundance estimates
rely on the effect of line blanketing by metals on photometric
bandpasses, but the strength of lines is also dependent on stellar temperature.
In the photometric plane defined by a line-blanketed index and a temperature
index, cooler stars may have $m_1$ or $hk$ indices similar to those for warmer,
more metal-rich stars. For this reason, an over correction for interstellar
reddening will produce a photometric abundance estimate that is too large.

The scheme we have used to determine and apply interstellar reddening
corrections is similar to that discussed by Anthony-Twarog \& Twarog
(1994)\markcite{a3} and Anthony-Twarog \& Twarog (1998)\markcite{a2}, in which
reddening estimates for red giants had to be obtained prior to the derivation
and revision of isometallicity lines in the $m_1$, $(b-y)$ and $hk$, $(b-y)$
diagrams.

Many, but not all, of the stars in the present sample are located
at high Galactic latitude, where one does not expect their indices to be
greatly affected by interstellar reddening.  We begin with an estimate of the
reddening E$(B-V)$ provided by the Burstein \& Heiles (1982)\markcite{bh} maps,
which correlate H I column densities to optical extinction by dust as implied
by counts of extragalactic sources. Estimates for most Galactic positions with
$|b| \geq 10 \arcdeg$ are available by examination of their published maps, or
a by use of a FORTRAN program which interpolates to provide E$(B-V)$ estimates.
For stars with latitudes above $|b| = 30 \arcdeg$, these estimates were applied
in full to the star's indices in the following amounts: E$(b-y) =$0.74
E$(B-V)$, $A_V =$4.3 E$(b-y)$, E$(c_1) =$0.2 E$(b-y)$, E$(m_1) =
-0.32$E$(b-y)$, and E$(hk) = -0.16$E$(b-y)$.  These reddening relations are
essentially those of Crawford \& Mandwewala (1976)\markcite{cm} with the $hk$
reddening discussed in Anthony-Twarog et al. (1991)\markcite{a1}.

Stars initially classified as red giants were also examined by an iterative
program that estimates metal abundance and the absolute magnitude implied for a
given unreddened color estimate.  This approach, discussed in Anthony-Twarog \&
Twarog (1994, 1998)\markcite{a3}\markcite{a2},
provides a distance estimate for red giants with $-0.8 \le {\rm [Fe/H]} \le
-2.4$, based on the $M_V, (B-V)$ relations for giant branches compiled by
Norris, Bessell, \& Pickles (1985)\markcite{nbp}.  If a red giant is nearer
than the edge of the Galactic dust disk, $d = 125$ pc/$\sin b $, its reddening
is pro-rated in the same proportion as its distance relative to $d$.   This
correction was applied to a handful of red giants with latitudes between
$10\arcdeg$ and $30\arcdeg$ above or below the Galactic plane.

A further check on the red giants' reddening is provided by the use of the
$c_0$, $(b-y)_0$ plane which, for giants with abundances below [Fe/H] $= -1.5$,
is relatively insensitive to metallicity differences.  Most stars classified as
dwarfs were corrected by the full amount of reddening implied by their Galactic
position, as were stars which appear to be horizontal branch stars, as it is
not possible to place these evolved stars unambiguously in the $c_1$, $(b-y)$
plane.  Exceptions are noted in Tables 5a--5c, which summarize unreddened
indices and photometric abundance estimates for three classes of stars (see
below). 

\section{Metallicity Estimates}

Beers et al. (1999)\markcite{bvir} describe the most recent calibration of
their techniques for obtaining metallicity estimates of stars based on
medium-resolution (1-2 \AA\ ) spectra and either measured or estimated
broadband $(B-V)_0$ colors.  They demonstrate that a combination of methods
based on the variation of an optimized index which measures the strength of the
CaII K line, and an Auto-Correlation Function (ACF) of the stellar spectrum,
can produce abundance estimates which have external errors on the order of
0.15-0.20 dex over the complete range of metallicities expected in the stellar
populations of the Galaxy, $-4.0 \le {\rm [Fe/H]} \le 0.0$.  As Beers et al.
discuss, there are a number of iterative steps which are applied in order to
arrive at a final estimate of stellar metallicity.  In the case of the present
application, we do not yet have measured ACFs for all of the stellar spectra of
our program stars with available spectroscopy, so we base our spectroscopic
estimates solely on the strength of the optimized CaII K index, which Beers et
al refer to as $KP$.  The estimate we obtain, [Fe/H]$_{K3}$ (adopting the Beers
et al.  nomenclature), is expected to be close to, but not exactly the same as,
the metallicity estimate which will be obtained once the full set of ACFs is
available.

Photometric estimates of stellar abundance may be derived by comparison of
reddening-corrected indices to calibrated isometallicity relations, or by means
of calibrations obtained with respect to stars with spectroscopically measured
metal abundance.  Because of the wide range of stellar types included in the HK
survey sample, a number of techniques have been employed.  Most of the stars in
our photometric sample have sufficient information to base an abundance
estimate on the $m_1$ as well as $hk$ index.  Some stars, however, have colors
or indices that place them outside the boundaries of the photometric
standardization or one of the metallicity calibrations.  Stars that are 
either too blue ($(b-y)_0 \leq 0.25$) or too red ($(b-y)_0 \geq 0.65$ 
for dwarfs, $(b-y)_0 \geq 0.90$ for giants) cannot be adequately treated 
by any of the photometric calibration schemes.

Which photometric index should work best for a given star?  An overview of the
complementary domains and sensitivities of the $m_1$ and $hk$ indices can be
gained by examination of indices generated by Soon et al. (1993)\markcite{soon}
from model atmospheres for dwarfs and giants over a broad range of color.
Figure 7 from their paper displays synthetic isometallicity curves for both
index planes.  Both indices, for example, exhibit diminished sensitivity for
blue stars but the problem is more critical for $m_1$ and is worse for dwarfs
than for giants.  Some serious degeneracies in the isometallicity curves are
predicted for the coolest stars, extending from high abundances over a range in
[Fe/H] that is broader for dwarfs than for giants.  There are regions in
temperature and [Fe/H] where the sensitivity of $m_1$ and $hk$ to [Fe/H]
variations are comparable, notably for metal-poor giants with $(b-y)$ colors
between 0.6 and 0.8.  For higher abundance dwarfs over a broad range of color,
$m_1$ displays finer [Fe/H] resolution than $hk$.

In addition to this rather rich palette of predicted sensitivities, the number
of extremely metal-poor stars with high dispersion spectroscopic abundances is
small, critically so below [Fe/H] = --2.5.  A final issue is demonstrated by
the relative positions of CD$ -38\arcdeg 245$ and BD $-18\arcdeg 5550$
in most photometric planes, inverted with respect to their well-measured but 
very low
[Fe/H] values;  this anomaly, commented on by Twarog \& Anthony-Twarog (1991)
and others, may be due to different abundances of nitrogen in these stars.
These various complications make a predetermination of a preferable photometric
metallicity index difficult for a heterogeneous population of stars.

\subsection{Spectroscopic Abundances}

Medium-resolution spectroscopy was obtained for the majority of our program
stars with the 2.1m telescope at KPNO, using the Goldcam spectrograph.  Details
of the acquisition of this data are discussed in detail in Beers et al.
(1999)\markcite{bvir}, and will not be repeated here.  Measurements of the $KP$
index were then used to obtain the metallicity estimate [Fe/H]$_{K3}$,
following the procedures outlined in Beers et al.

Since it is our plan to compare the abundances obtained from the photometric
and spectroscopic approaches, it was decided that the Str\"omgren $(b-y)_0$
colors for the program stars would be used to {\it predict} the appropriate
broadband $(B-V)_0$ color that is used as one of the inputs for the metallicity
estimation procedure, adopting the transformation used above, $(B-V)_0 =
1.35(b-y)_0$.  Beers et al. (1999)\markcite{bvir} discuss the means by which an
estimate of a reddening corrected broadband $(B-V)_0$ color may be obtained
from a correlation of the strength of the Balmer H$-\delta$ line based on
medium-resolution spectroscopy.  As a check, we have compared this estimated
color with the predicted broadband colors obtained from the Str\"omgren data.
For a number of stars, the agreement was less than ideal (differing from one
another by more than 0.1 magnitudes), which might have a significant effect on
the derived metallicity estimates.  In Tables 5a--5c below, we have indicated
the cases where this color discrepancy exists by appending a ``:'' following
the derived spectroscopic abundance estimate.

\subsection{Red Giants}

Although the stars classified as giants are a minority ($\leq 15$\% of the
original photometric survey) in our sample, they have the advantage that two
independent photometric calibrations can be applied which are securely rooted
in homogeneous scales of spectroscopic abundances.  A calibration using $m_1$
indices was derived by Anthony-Twarog \& Twarog (1994)\markcite{a3} as part of
a comprehensive effort to use Str\"omgren photometry to ascertain reddenings,
distances, and metal abundance estimates for a sample of red giants.  Self
consistent estimates for all of these quantities are necessary steps in the
development of a photometric calibration.  Distances affect the foreground
reddening estimates.  Reddenings, in turn, affect the photometric metallicity
estimates in the sense that larger reddening corrections generally lead to
larger estimates of [Fe/H].  In Anthony-Twarog \& Twarog (1994)\markcite{a3},
isometallicity lines in the $m_0$, $(b-y)_0$ plane were derived for 58 giants
with abundances in the range $-2.7 \le {\rm [Fe/H]} \le -1.0$.  The photometry
is consistent with the photometric system of Bond (1980)\markcite{b1} and the
abundances are derived, as in Twarog \& Anthony-Twarog (1991)\markcite{t2},
from a variety of high-dispersion spectroscopic studies mapped to a common
system based on Gratton \& Sneden (1988)\markcite{g88}.

The same precepts were followed by Anthony-Twarog \& Twarog (1998)\markcite{a2}
to develop isometallicity lines in the $hk_0$, $(b-y)_0$ plane.   The catalog
of homogeneously merged spectroscopic [Fe/H] values was enlarged by addition of
studies by Sneden et al. (1991)\markcite{s91} and Kraft et al.
(1992)\markcite{k92}, and others, as described in Anthony-Twarog \& Twarog
(1998)\markcite{a2}.  This calibrated plane yields [Fe/H] estimates for
red giants more metal-poor than [Fe/H]$ = -1.0$ that are precise to $\pm 0.2$
dex.  The performance of the index declines for higher abundance giants over a
wide range of color, as well as for the bluest stars.  Satisfactory estimates
of abundance for giants bluer than $(b-y)_0 = 0.3$ or redder than $(b-y)_0 =
0.9$ cannot be obtained from this approach.

Final photometric estimates of [Fe/H] were obtained for each of our
program stars identified as likely red giants using their colors, $m_0$ and
$hk_0$ indices by interpolation between the isometallicity relations published
by Anthony-Twarog \& Twarog (1994)\markcite{a3} and Anthony-Twarog
\& Twarog (1998)\markcite{a2}.  The errors associated with each abundance 
estimate are based on the photometric error listed in Table 3, and the slope 
of the the isometallicity relationship.   The contribution to the abundance
error must, in principle, include the contribution of $\sigma_{by}$ as well as
the error in $m_0$ or $hk_0$.  A contribution was added in quadrature to
account for the typical uncertainty in abundance associated with each
calibration.  These contributions to our estimated errors are 0.16 dex for
abundances based on $m_1$, and 0.20 dex for the $hk$-based abundances.  We
believe these errors to be conservative upper limits to the systematic error
for each photometrically derived abundance.

Results for 63 program stars classified as red giants are presented in Table
5a.  In this table, the first column lists the star identification.  Columns
(2)--(6) list the reddening-corrected photometric quantities for each
star.  The reddening estimate for each star, E$(b-y)$, is listed in column (7).
Photometric estimates of abundance based on the reddening-corrected $m_1$ index,
and its associated error, are listed in columns (8) and (9), respectively.
Columns (10) and (11) list photometric abundance estimates and errors
based on the reddening-corrected $hk$ index.  Columns (12) and (13) provide the
spectroscopic estimates of metal abundance, and associated errors, obtained
using the Beers et al. (1999)\markcite{bvir} calibration.  Metallicity
estimates for stars which fall outside [Fe/H] limits of the photometric 
calibrations are listed in the table as upper or lower limits.  In cases
where a photometric abundance estimate is missing altogether, it is because
the associated error in [Fe/H] implied by the photometry is larger than 1.0
dex.

Figure 2 shows a comparison of estimated abundances for the red giants with
acceptable metallicity determinations from Table 5a.  The one-to-one lines
in the figures have endpoints that reflect the formal limits of the photometric
calibrations, $-3.4 \le {\rm[Fe/H]}_{hk} \le -0.5$ for the $hk$-based
calibration, and $-2.7 \le {\rm [Fe/H]}_{m1} \le -1.0$ for the calibration based
on $m_1$.  These limits may explain, in part, the general tendency for the
photometric abundances to appear too low for stars of moderate metal abundance
([Fe/H] between --1.0 and --0.4) and for the apparent decline in sensitivity of
the $m_1$-based abundances below [Fe/H]$= -2.5$.

It should be noted that the two photometric abundance estimates are not
entirely independent, as both are based on the same $(b-y)$ color.
Furthermore, the high dispersion spectroscopic abundances upon which the two
metallicity calibrations are based are essentially identical.  The scatter
between abundance estimates based on $m_1$ and on $hk$ is approximately 0.35
dex.  A slightly larger scatter, $\sim 0.40$ dex, is found between the
photometric indices and the spectroscopic estimates.  This level of
scatter between the different methods is in accord with our expectations based
on the individual errors obtained for each star.

\subsection{Main Sequence Stars}

The largest portion of our program sample ($\sim 55$\% of the original sample) 
is comprised of stars
classified as main sequence dwarfs based on their color and surface gravity
indices.  The primary calibration used to derive abundance estimates for F and G
dwarfs from $m_1$ has been adopted from Schuster \& Nissen (1989)\markcite{s1}.
The F-star calibration includes a term involving surface gravity through the
$c_1$ index; both calibration equations incorporate several terms expressing
linear and higher-order dependence on $m_0$ and $(b-y)_0$.  The calibrations
are limited to [Fe/H] values between --3.5 and +0.2 for the F dwarfs, and
between --2.6 and +0.4 for the G dwarfs.  The spectroscopic basis for these
calibrations is consistent with a scale that yields an abundance [Fe/H]$ =
+0.13$ for the stars in the Hyades open cluster.

Dwarfs were part of the original sample of stars used to define the $hk$ index
by Anthony-Twarog et al. (1991)\markcite{a1}, but a thorough analysis of dwarfs
for the purpose of deriving a metallicity calibration was not attempted at that
time.  A preliminary calibration was prepared for this purpose, based on
photometric data for several dozen dwarfs with photometry in the Twarog \&
Anthony-Twarog (1995)\markcite{t1} catalog and with available high-dispersion
spectroscopic abundances in the homogeneous catalog compiled by Anthony-Twarog
\& Twarog (1998)\markcite{a2} for the red giant calibration.  It was apparent
at the time that this sample of ~50 stars was spread too thinly in abundance
and color to provide an adequate outline of the $hk, (b-y)$ isometallicity
lines. In particular, very few dwarfs with abundances between [Fe/H] $= -1.0$
and the abundance of the Hyades were available.  Thus, the dwarf abundance
calibration effort was postponed.

With access to the spectroscopic abundances for the nearly 300 dwarfs in our
photometric sample based on the Beers et al. (1999)\markcite{bvir} calibration,
isometallicity lines were re-drawn using the [Fe/H]$_{K3}$ abundances as a
homogeneous set of spectroscopic estimates.  It must be noted, therefore, that
ensuing comparisons between the $hk$-based photometric abundances for dwarfs
and the [Fe/H]$_{K3}$ estimates are not independent.  Provisional 
isometallicity lines
for main sequence stars with $0.25 \le (b-y)_0 \le 0.45$ are drawn in Figure 3.
Superposed on the isometallicity grid are the colors for 51 field dwarfs from
the 1995 catalog with different symbols distinguishing stars with [Fe/H] values
as high as --1.0.  Three field stars with high dispersion spectroscopic
abundances below [Fe/H $= -2.5$ are noted in the figure:  in order of
increasing color these are BD$+1\arcdeg 2341$, G 64-12, and CS 22876-0032.  The
adopted spectroscopic abundances for these three field stars are --2.50, --3.01
and --3.78, respectively.  A comparison of abundances for these 50 stars
derived from the isometallicity lines with the spectroscopic values has a
standard deviation of 0.27 dex.

With the reminder that the photometric abundance estimates based on $m_1$ and
on $hk$ are still linked through their common color, and that the dwarf
abundances based on $hk$ are now tied to the K3 spectroscopic abundances, Table
5b summarizes the reddening-corrected photometry information and abundances for
stars with derivable abundances and associated errors, with column headings
as for Table 5a.  Schuster \& Nissen (1989)\markcite{s1}
estimate a statistical uncertainty of 0.16 dex for [Fe/H] estimates based on
$m_1$ for F and G dwarfs, which we have added in quadrature to the photometric
error on the abundance estimate for the $m_1$-derived abundances listed in
Table 5b.  Errors for the abundances based on $hk$ indices include a quadratic
contribution of 0.27 dex.

Figure 4 shows a comparison of the photometric and spectroscopic abundances
detailed in Table 5b.   In each panel, a line of unit slope has been drawn.
The interdependence of the $hk$ calibration on the [Fe/H]$_{K3}$ abundance
scale is apparent in the lowest panel of Figure 4, although the 
correlation between these two indicators appears to soften at the most 
metal poor end.   The dispersion about the line for 222 stars in common 
is approximately 0.4 dex.  The upper panel of Figure 4
illustrates a comparison of comparable dispersion between the two photometric
indicators for 162 stars in common.  The observed correlation between the
[Fe/H]$_{K3}$ abundances and those based on $m_1$ is reassuring, although the
scatter appears somewhat larger.

Of the 289 stars with photometric abundances listed in Table 5b, 19 stars have
$hk_0$ less than 0.2 and/or $m_0$ less than 0.0.  Such low indices are
fairly robust indicators of extreme metal paucity.  For most of these stars,
the low abundance limit of the $m_0$-based [Fe/H] calibration is violated.
For seven stars, upper limits of $-$2.5 or $-$3.0 provide the strongest limit
on [Fe/H] that the $hk_0$ calibration can provide.  Two stars exhibit clear
discrepancies between the various metal abundance indicators.  BS 17438-0031
has a low $m_0$ index, too low for the photometric calibration to supply an
[Fe/H] estimate, but the $hk_0$ estimate of metallicity is consistent with the
[Fe/H]$_{K3}$ abundance estimate.  The low $m_1$ index for another star, BS
16924-0026, implies a low photometric metallicity estimate that 
is consistent with
the [Fe/H]$_{K3}$ abundance, yet the $hk$ index indicates a higher abundance.
With the exclusion of these two stars, the $hk$-derived and the [Fe/H]$_{K3}$
abundances are consistent, with a scatter between them of roughly 0.45 dex, in
identifying these 19 stars as extremely metal-poor.  Four stars, BS 16922-0090,
BS 17139-0018, BS 17448-0021 and CS 22189-0003, have photometric estimates of
[Fe/H] running from $-$2.4 to below $-$3.0, but have no available spectroscopic
abundance estimates.

\subsection{Field Horizontal-Branch Candidates}

A number of stars in our program sample are relatively blue ($(b-y)_0 \leq
0.40$) and have high $c_0$ indices, greater than 0.50 (in some cases,
greater than 1.0), indicative of highly evolved or early-type stars with
substantial foreground reddening.  Although we discuss stars covering a large
range in $c_0$ that lie within the color limits $0.2 \le (b-y)_0 \le 0.4$, more
stringent criteria have been proposed by Gratton (1998)\markcite{g98} for the
classification of field horizontal-branch (FHB) stars.  In addition to 
restrictions $m_0 \leq 0.15$, and $0.4 \le c_0 \le 0.8$, Gratton suggested that
FHB stars should lie within a band in the pseudo-HR diagram extending upwards
by 0.3 magnitudes in $c_0$ from a line joining $(b-y)_0 = 0.2$, $c_0 = 0.7$ to
$(b-y)_0 = 0.4$, $c_0 = 0.3$.  Stars meeting this tighter criterion are
considered candidate FHB stars in the following.

While there is not a calibration in place for horizontal-branch stars in the
$hk$ system, we have adapted a calibration derived by Baird
(1999)\markcite{bai} based on spectroscopic and photometric observations of RR
Lyrae stars of known metallicity.  Baird's work demonstrated the relative phase
invariance of loci traced by RR Lyrae's in the $hk_0, (b-y)_0$ diagram.  This
calibration is shown in Figure 5 by dashed lines in the region appropriate for
FHB stars.  The calibration lines from Anthony-Twarog \& Twarog
(1998)\markcite{a2} for red giants are also shown in this diagram along with
red giants from the present study.

Table 5c summarizes photometric abundances derived from the position 
in the $hk_0$,$(b-y)_0$ plane of our program stars classified as FHB 
and other high-$c_0$ stars.   Due to the
limited range in [Fe/H] for this preliminary calibration, a number of stars are
listed with only upper or lower limits for [Fe/H].  The table is segregated
into groups conforming to Gratton's suggested photometric criteria for FHB
stars (the first 25 stars) and an additional 23 stars with 
generally higher $c_0$ indices.

\section{Conclusions}

A subset of the extensive HK survey of candidate metal-poor stars has been
studied with extended Str\"omgren photometry and medium-resolution
spectroscopy.  Photometry for 521 candidates stars formed the basis for
analysis of foreground reddenings, luminosity classification and metallicity
estimates based on $m_1$ and $hk$ indices.  Although a number of stars were
outside calibrated color and index limits, and others would exceed the
metallicity limits of the photometric calibrations, photometric abundances for
63 red giants, 289 dwarfs and 48 blue horizontal branch stars have been
estimated and compared to abundances based on spectroscopic analysis of the Ca
II K line.

Among the red giants, nearly a third of the candidates appear to have metal
abundances below [Fe/H] = --2.5 and a handful may have abundances below
[Fe/H] = --3.0.  This class of stars is favored with two well-developed
photometric calibrations for metallicity based on both $m_1$ and $hk$.  A
preliminary calibration developed by Baird (1999)\markcite{bai} for RR Lyrae
stars has been applied to the $hk$ indices and $(b-y)$ colors for the FHB
candidate stars, identifying two of these candidates as having abundances
below [Fe/H] = --3.0.

Although photometric calibrations for main sequence stars based on $m_1$ are
well established, a preliminary calibration for metal-poor dwarfs based on $hk$
indices was developed for the present investigation that is tied to the
medium-resolution abundances based on the K line. In part because the
photometric calibrations do not extend to uniformly extreme low values of
[Fe/H], the yield of extreme candidates from this unevolved population is
somewhat lower, with approximately 30 stars having one or both photometric
abundances below [Fe/H] = --2.5.  Approximately 20 stars have photometric
indices indicative of more extreme metal-paucity.

We close by presenting in Table 6 an abridged list of identifications,
equatorial coordinates, $V$ and $(b-y)$ along with luminosity classification,
an averaged photometric abundance estimate and the [Fe/H]$_{K3}$ abundance
estimate.  These 42 stars fit the following criteria:  any of the two or three
methods implying [Fe/H] $ \leq -3.0$; an estimate of [Fe/H] $\leq -2.5$ that is
the only abundance estimate available; two of three abundance estimates below
[Fe/H $ = -2.5$.

\acknowledgements

The support of the General Research Fund of the University of Kansas to BJAT
and BAT is gratefully acknowledged.  Some of this research was supported by a
grant from the National Science Foundation to the University of Kansas and to
Kansas State University through its EPSCoR program.  TCB acknowledges partial
support for this research from grants AST 92-22436 and AST 95-29454 from the
National Science Foundation.  All of us would like to thank the staff of Kitt
Peak National Observatory for their support of these observations, particularly
Ed Carder.

\newpage
\figcaption{Instrumental errors for photometric colors and indices for
521 program stars, as a function of $V$ magnitude.  \label{fig1}}
\figcaption{Intercomparison of [Fe/H] determinations for red giants
by spectroscopic methods (K3) and photometric methods, based on $m_1$ and 
$hk$ indices.\label{fig2} }
\figcaption{Preliminary isometallicity lines in the $hk_0$,$(b-y)_0$ plane
for main sequence stars.  The abundance scale is tied to the K3
abundances for the program stars discussed in this paper.  A number of
field dwarfs with high dispersion spectroscopic abundances are projected
on the isometallicity grid for comparison.  \label{fig3} }
\figcaption{Intercomparison of [Fe/H] determinations for main sequence stars 
by spectroscopic methods (K3) and photometric methods, based on $m_1$ and 
$hk$ indices.\label{fig4}}
\figcaption{Candidate red giants, horizontal branch stars and other blue
stars with high $c_0$.  Filled squares indicate stars which satisfy 
color and surface gravity criteria as horizontal branch stars; triangles
are used for stars classified as red giants.  The calibration of Anthony-Twarog
\& Twarog (1998)\markcite{a2} is shown for the red giants; a preliminary
calibration developed by Baird (1999)\markcite{bai} for RR Lyrae stars, is
employed for the bluer stars.\label{fig5}
} 
\begin{table}
\dummytable\label{tbl$-$1}
\dummytable\label{tbl$-$2}
\dummytable\label{tbl$-$3}
\dummytable\label{tbl$-$4}
\dummytable\label{tbl$-$5}
\dummytable\label{tbl$-$5a}
\dummytable\label{tbl$-$5b}
\dummytable\label{tbl$-$5c}
\dummytable\label{tbl6}
\end{table}

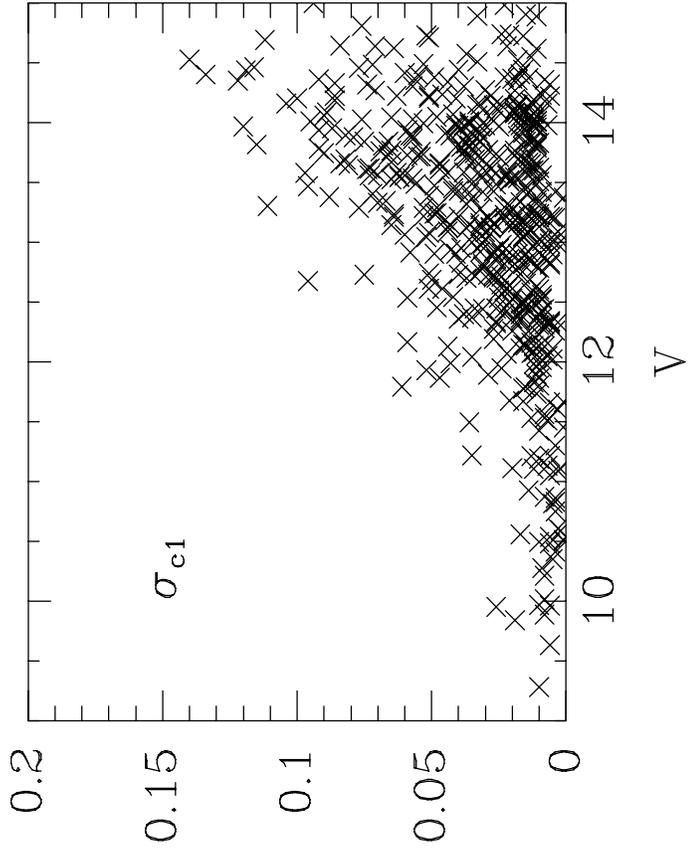
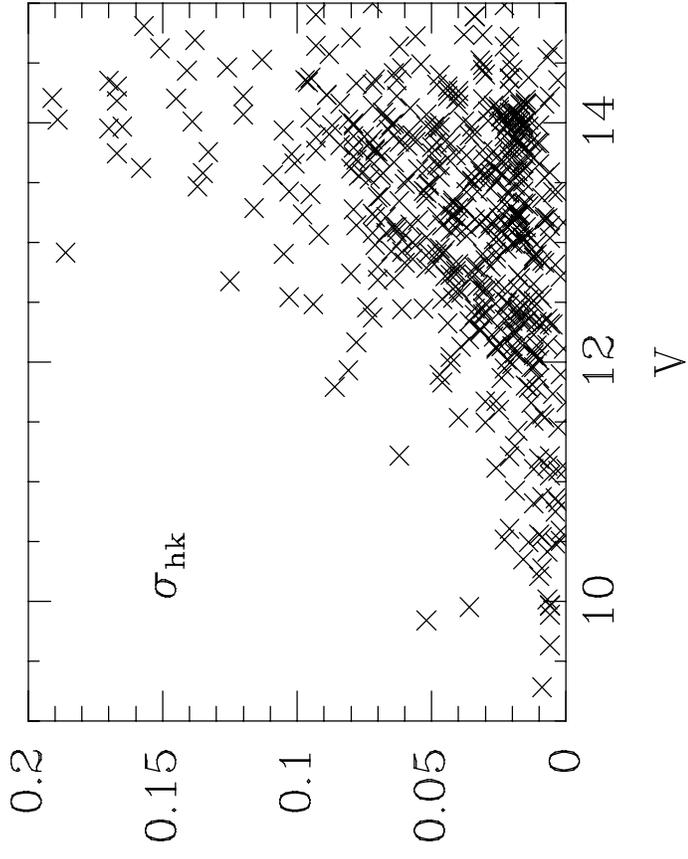
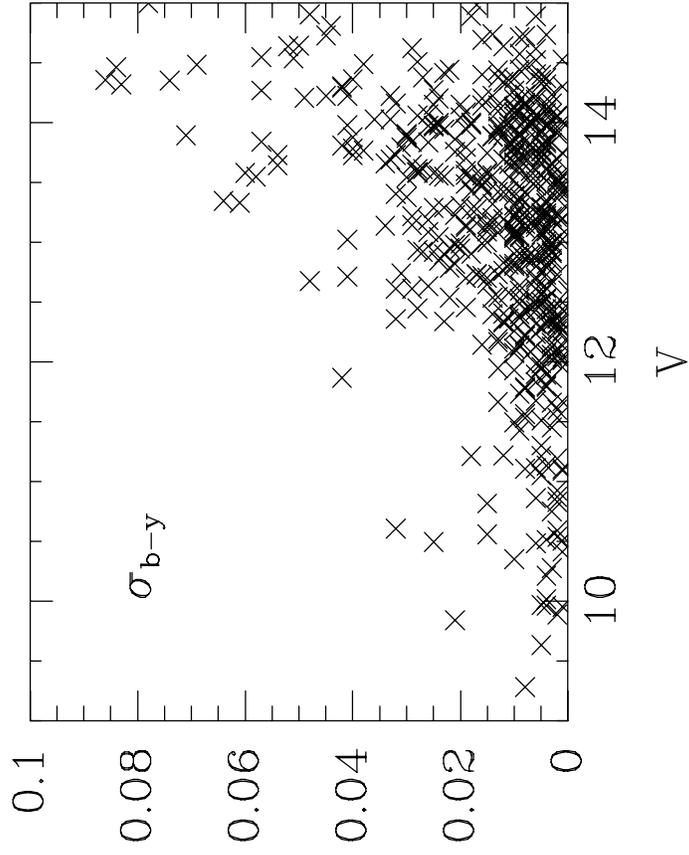
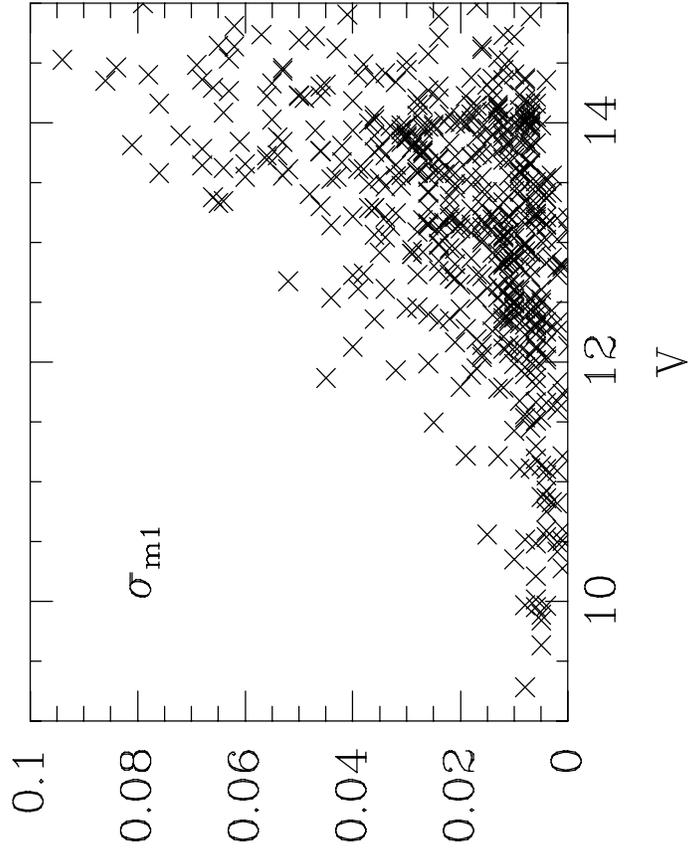
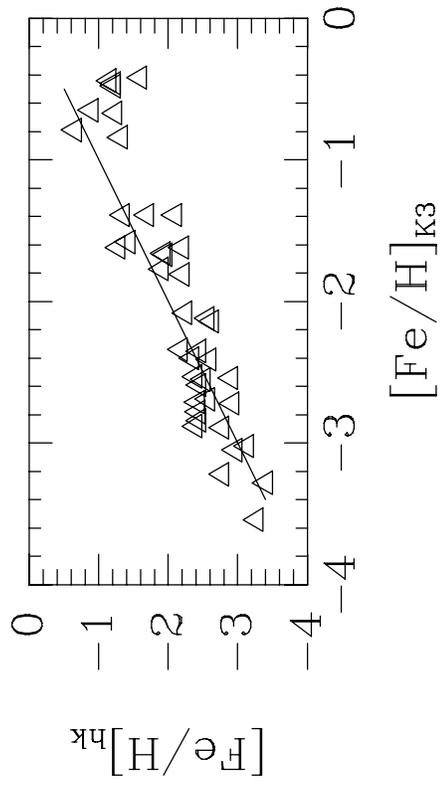
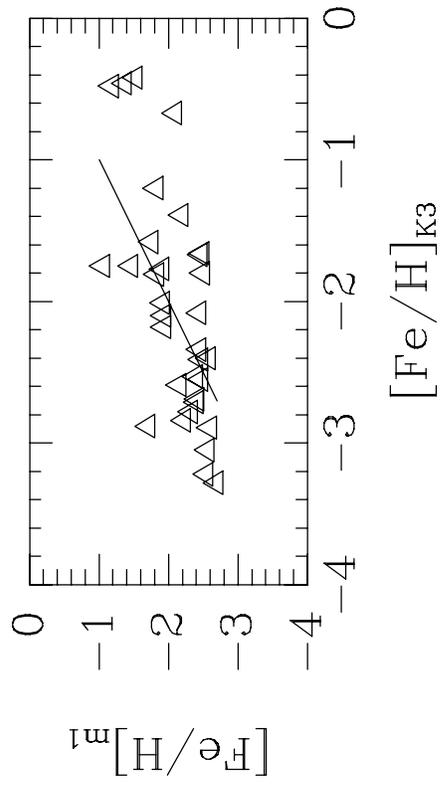
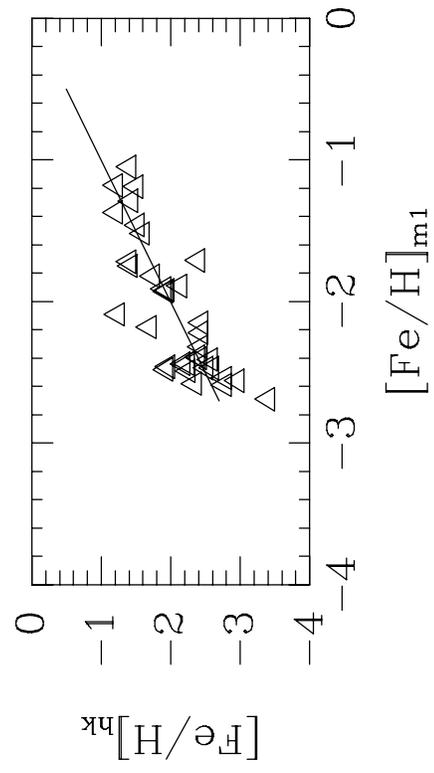
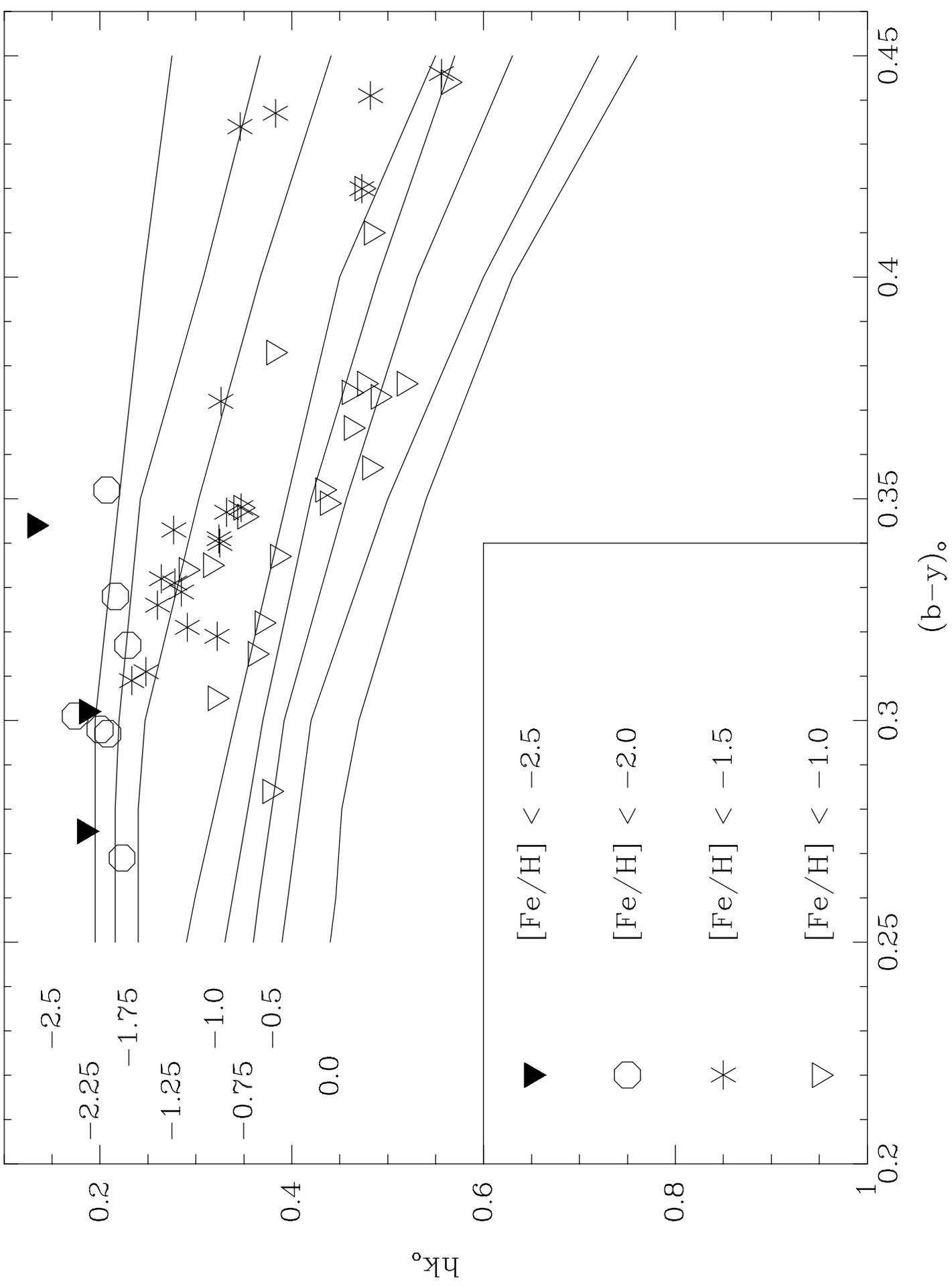
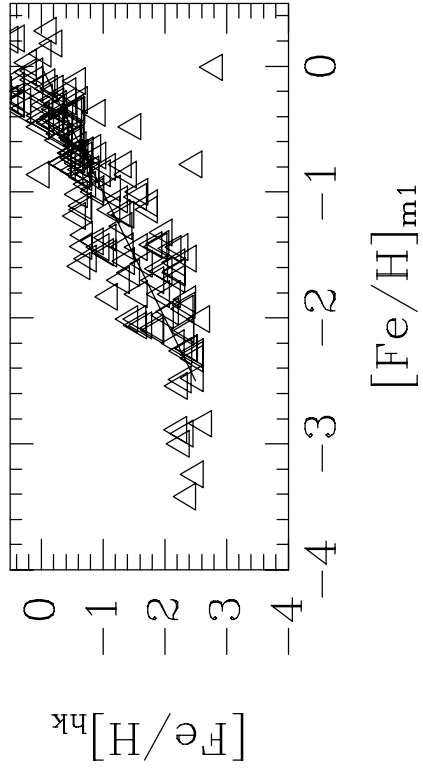
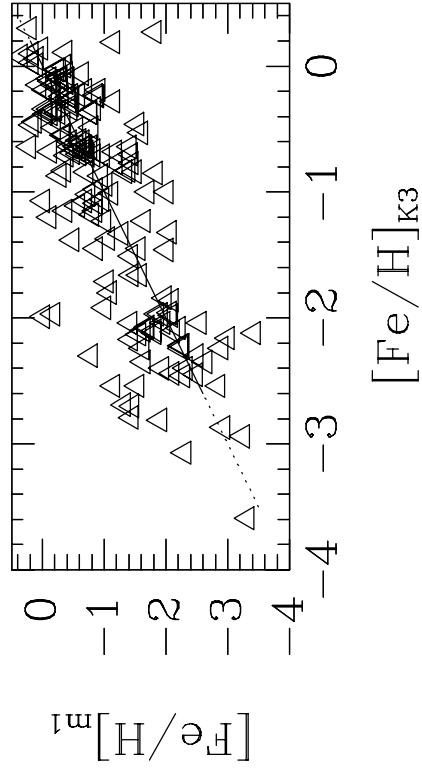
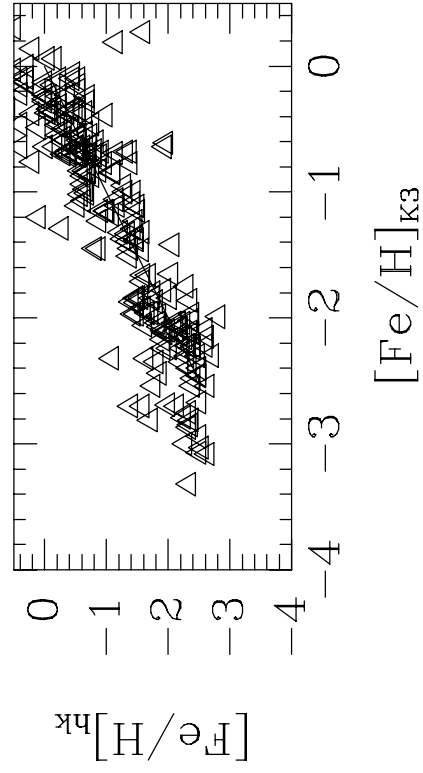
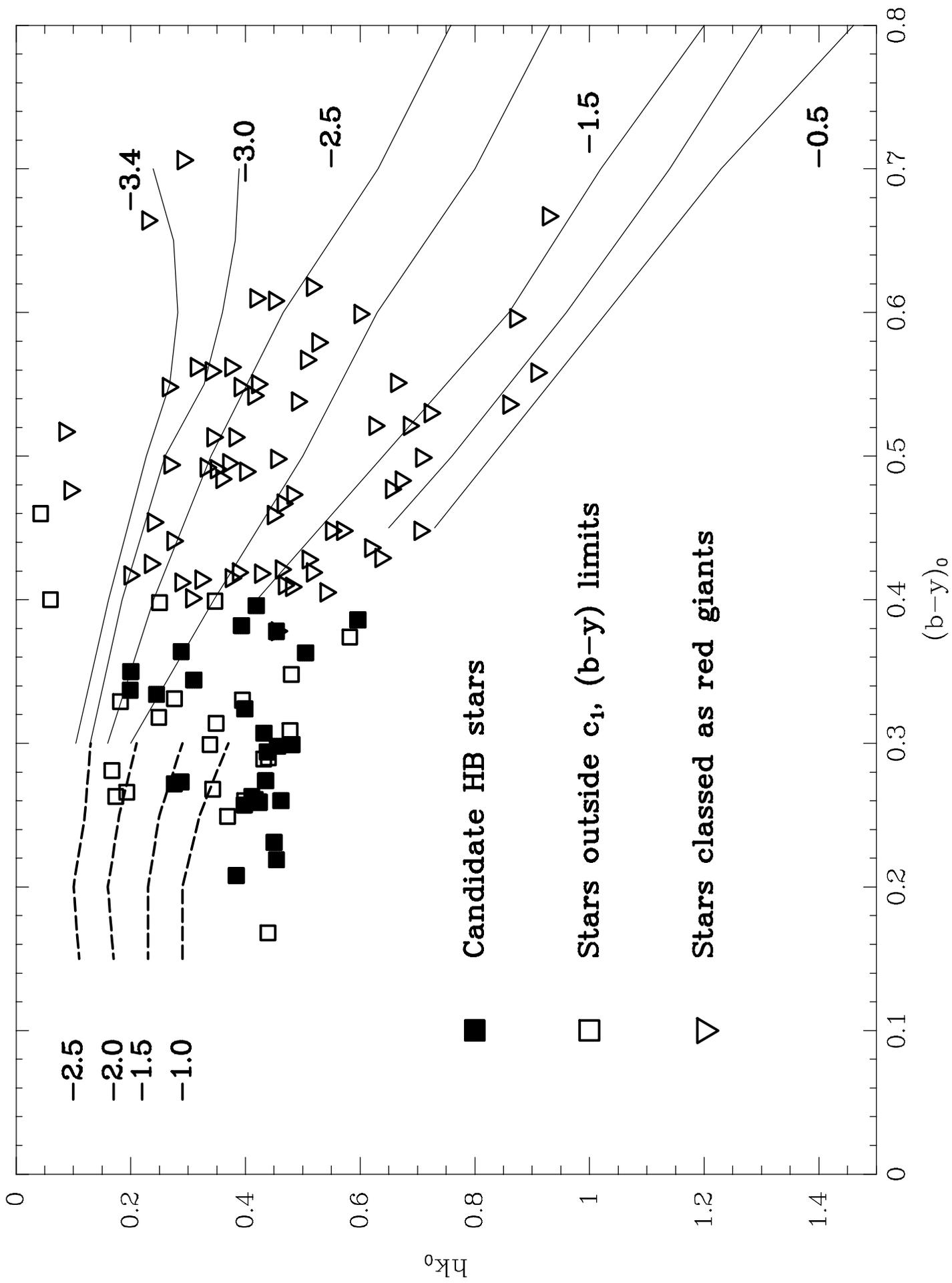
\end{document}